\documentclass[reprint, superscriptaddress, prl]{revtex4-2}

\usepackage{graphicx}
\usepackage{dcolumn}
\usepackage{bm}
\usepackage{color}
\usepackage[colorlinks, linkcolor=blue, anchorcolor=blue, citecolor=blue]{hyperref}
\usepackage{siunitx}
\DeclareSIUnit\nounit{{}}
\usepackage[normalem]{ulem}
\usepackage{comment}
\usepackage[utf8]{inputenc}

\begin{document}

\title{How liquid-liquid phase separation induces active spreading}
\author{Youchuang Chao}
\email{youchuang.chao@ds.mpg.de}
\author{Olinka Ramírez-Soto}           
\author{Christian Bahr} 
\author{Stefan Karpitschka}
\email{stefan.karpitschka@ds.mpg.de}
\affiliation{Max Planck Institute for Dynamics and Self-Organization, 37077 G{\"o}ttingen, Germany}

\date{\today}

\begin{abstract}
The interplay between phase separation and wetting of multicomponent mixtures is ubiquitous in nature and technology and recently gained significant attention across scientific disciplines, due to the discovery of biomolecular condensates.
It is well understood that sessile droplets, undergoing phase separation in a static wetting configuration, exhibit microdroplet nucleation at their contact lines, forming an oil ring during later stages.
However, very little is known about the dynamic counterpart, when phase separation occurs in a non-equilibrium wetting configuration, i.e., spreading droplets.
Here we report that liquid-liquid phase separation strongly couples to the spreading motion of three-phase contact lines.
Thus, the classical Cox-Voinov law is not applicable anymore, because phase separation adds an active spreading force beyond the capillary driving.
Intriguingly, we observe that spreading starts well before any visible nucleation of microdroplets in the main droplet.
Using high-speed ellipsometry, we further demonstrate that surface forces cause an even earlier nucleation in the wetting precursor film around the droplet, initiating the observed wetting transition.
We expect our findings to enrich the fundamental understanding of phase separation processes that involve dynamical contact lines and/or surface forces, with implications in a wide range of applications, from oil recovery or inkjet printing to material synthesis and biomolecular condensates.
\end{abstract}

\maketitle


\textit{Introduction.---}Phase separation or demixing of homogeneous liquid mixtures into two or more distinct phases frequently occurs in nature and everyday life, and critically impacts a variety of engineering applications~\citep{lohse2020physicochemical}, such as oil recovery~\citep{muggeridge2014recovery}, inkjet printing~\citep{xia2005controlled}, and materials synthesis~\citep{chao2020emerging}.
In most practical situations, phase separation processes occur in heterogeneous environments, i.e., in contact with surfaces, because the interaction with surfaces facilitates nucleation~\citep{turnbull1950kinetics}. Thus the interplay of phase separation and wetting is often nontrivial and can not be ignored~\citep{gelb1999phase, tanaka2001interplay}.
For instance, the wettability of rock surfaces can strongly affect separation efficiency of the crude oil-water mixture for recovering oil from underground reservoirs~\citep{muggeridge2014recovery}.
Even for a single-component liquid, the kinetics of phase transition between different liquid states can be altered by the presence of solid surfaces~\citep{murata2010surface, murata2012liquid}.
In addition to technical applications, coexistence of phase separation and wetting is found in biological settings. 
A typical example is the protein condensation, a key process for living cells to form membraneless organelles~\citep{banani2017biomolecular, shin2017liquid}, which happens not only in bulk cytoplasm, but frequently on surfaces like the nucleus, microtubuli, and lipid bilayers \citep{brangwynne2015polymer, ditlev2021membrane}.
In the latter case, the wetting properties of biological membranes plays an essential role, for instance, in regulating autophagy of the phase-separated compartments~\citep{agudo2021wetting}. 

Earlier studies have revealed the physical mechanisms of heterogeneous nucleation, such as the effect of particle sizes and surface properties on the nucleation efficiency~\citep{fletcher1958size, kim2008kinetics}, or the impact of preferential wetting on spinodal decomposition in binary liquid mixtures~\citep{wiltzius1991domain, genzer1997wetting, tanaka2001interplay, puri2005surface}.
Recent work has demonstrated how selective evaporation of sessile droplets in various static wetting configurations modulates nucleation and coarsening processes, e.g., in evaporating ternary Ouzo drops~\citep{tan2016evaporation, tan2017self}.
Most of those studies focus on the aspect of static wetting, i.e., pinned three-phase contact line conditions~\citep{fletcher1958size, genzer1997wetting, kim2008kinetics, wiltzius1991domain, puri2005surface, li2011membrane, tan2016evaporation, tan2017self, jensen2015wetting}.
Therefore, it remains unclear how the dynamic counterpart, e.g., moving contact lines~\citep{de1985wetting, bonn2009wetting, snoeijer2013moving} interact with phase separation, despite its abundance in many natural and industrial scenarios.

\begin{figure}
\centering
\includegraphics[width=8.5cm]{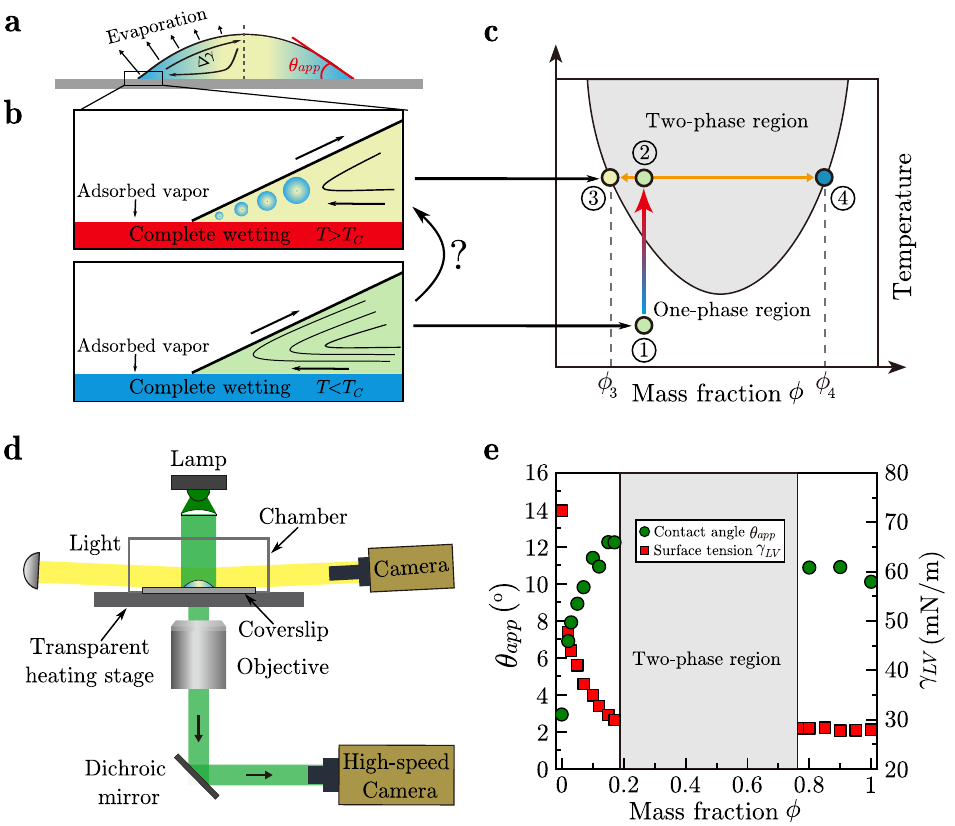}
\caption{Experimental system and setup.
(a)~Schematic cross-section of an evaporating binary droplet on complete wetting surfaces, with a non-zero apparent contact angle due to solutal Marangoni flows.
(b)~Zoom to the contact line region before and during phase separation, triggered by heating the substrate.
(c)~Schematic phase diagram of a binary-liquid system with a lower critical solution temperature (LCST). Upon heating from the one-phase region \raisebox{.5pt}{\textcircled{\raisebox{-.9pt} {1}}} to two-phase region \raisebox{.5pt}{\textcircled{\raisebox{-.9pt} {2}}}, the mixture separates into two phases \raisebox{.5pt}{\textcircled{\raisebox{-.9pt} {3}}} and \raisebox{.5pt}{\textcircled{\raisebox{-.9pt} {4}}}.
(d)~Schematic of the experimental setup, with simultaneous observation from the side and below.
(e)~Apparent contact angle $\theta_{app}$ (green) and surface tension $\gamma_{LV}$ (red) of the DPnP-water mixture versus DPnP mass fraction $\phi$ on complete wetting surfaces at $T \sim 20\, ^\mathrm{o}$C. 
}
\label{fig:PhaseDiagram}
\end{figure}

Here we explore the interplay between phase separation and wetting dynamics, using droplets of an evaporating, non-ideal, binary liquid mixture with a well-defined miscibility gap on complete wetting substrates.
We adopt droplets of water and glycol ethers as a model system that exhibits a lower critical solution temperature (LCST) close to room temperature (see Supplementary Information, Fig.~S1 and Table~S1).
In the one-phase region, due to solutal Marangoni flows, the droplet maintains a quasi-stationary, contracted state with a nonzero apparent contact angle $\theta_{app}$ and a high mobility, i.e., an unpinned contact line~\citep{cira2015vapour, karpitschka2017marangoni, malinowski2020nonmonotonic, molina2021droplet, parimalanathan2021controlling, williams2021spreading, wu2021contact, charlier2022water} (Fig.~\ref{fig:PhaseDiagram}a).
We then trigger phase separation by driving the droplet into the miscibility gap, by heating and/or selective evaporation (Fig.~\ref{fig:PhaseDiagram}b,c).
Surprisingly, upon liquid-liquid phase separation, we see actively driven droplet spreading.
Notably, spreading occurs before any visible nucleation of microdroplets in the main drop.
High-speed ellipsometry reveals an even earlier phase separation in the precursor film around the droplet, evidencing the strong coupling of phase separation and surface forces, which leads to the observed wetting transition.

\textit{Experimental setup and system.---}
To heat droplets in a precise manner, we built a computer-controlled heating system, which was mounted on top of an inverted microscope (Nikon Eclipse Ti2E).
As substrates, we used precision microscopy coverslips (VWR, thickness 0.17 mm) and one-side frosted microscopy slides (Corning,  thickness 0.96 $\sim$ 1.06 mm), cleaned by piranha solution or plasma treatment to generate complete wetting surfaces, or by ethanol for partially wetted surfaces.
Droplets of initial volumes $\Omega = $ 0.5--2 $\SI{}{\micro\liter}$ were deposited onto the substrates and then heated at a controlled rate.
Bottom-view and side-view images were simultaneously recorded by a high-speed camera (Phantom VEO 4K-L, 50 -- 500 fps) on the microscope and a CMOS camera (Point Grey Grasshopper2, 27 fps) attached to a telecentric lens, respectively (Fig.~\ref{fig:PhaseDiagram}d).
We used binary mixtures of water (``Milli-Q", resistivity 18 M$\Omega$ cm) and di(propylene glycol) propyl ether (DPnP, Sigma-Aldrich, $\ge$ 98.5\%, mass fraction $\phi$), unless stated otherwise (see Methods for more experimental details). 
Fig.~\ref{fig:PhaseDiagram}(e) shows apparent contact angle $\theta_{app}$ (green) and surface tension $\gamma_{LV}$ (red) versus $\phi$ of our DPnP-water mixture in the one-phase region on completely wetted substrates.
However, the apparent contact angles are quasi-statically non-zero: the single-phase water-rich binary mixture exhibits strong Marangoni-contraction, whereas the glycol ether-rich mixture shows autophobing \cite{hack2021wetting}.
What happens when the droplet is now forced into the two-phase region (Fig.~\ref{fig:PhaseDiagram}b,c)?

\begin{figure*}
\centering
\includegraphics[width=15cm]{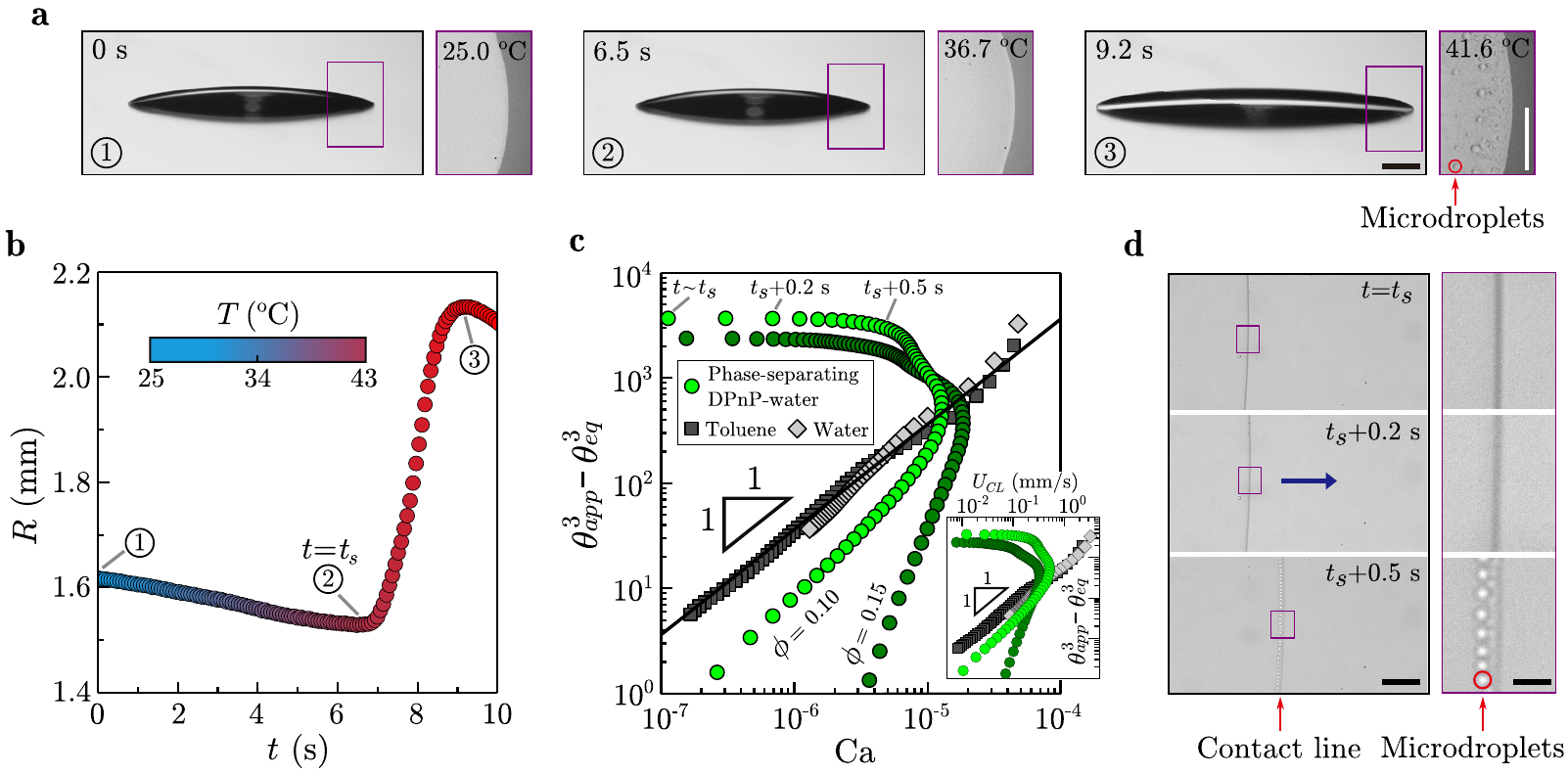}
\caption{Active spreading of a $\SI{1}{\micro\liter}$ phase-separating DPnP-water droplet on a substrate heated at $\sim 1.8\, ^\mathrm{o}$C/s.
(a) Image sequence of the side aspect of the droplet ($\phi = 0.1$) (left) together with the bottom view of the contact line region (right). Scale bars are 0.25 mm.
(b) Droplet radius $R$ versus time $t$ and temperature $T$ as color code. $t_s$ marks the onset of spreading.
(c) $\theta_{app}^3 - \theta_{eq}^3$ versus the capillary number $\mathrm{Ca}$ for phase-separating DPnP-water ($\phi = 0.1$ \& $0.15$, light \& dark green, respectively), compared to simple fluids toluene and water (dark and light grey, respectively). Simple fluids follow the Cox-Voinov law Eq.~(\ref{eq:CVLaw}), collapsing onto a single line of slope 1 (black line).
Inset: same data, as function of the physical $U_{CL}$ (also see Supplementary Information, Fig.~S3).
(d) High-resolution images of the contact line region. Spreading (direction indicated by the arrow) starts first, followed by nucleation and growth of microdroplets. Scale bars are $50$ and $\SI{10}{\micro\meter}$ for main and zoom panels, respectively.
}
\label{fig:MacroDropDynamics}
\end{figure*}

\textit{Abrupt spreading.---}We begin with investigating the macroscopic dynamics by heating droplets in pinned and unpinned situations, i.e., partially and completely wetted, respectively. 
Fig.~\ref{fig:MacroDropDynamics}(a) illustrates a typical image sequence of a $\SI{1}{\micro\liter}$ DPnP-water binary droplet with $\phi = 0.1$ heated on a completely wetted substrate.
During heating the substrate, we first observe enhanced contraction  (see Fig.~\ref{fig:MacroDropDynamics}a 0--6.5 s, Fig.~\ref{fig:MacroDropDynamics}b), owing to the increased selective evaporation and thus intensified Marangoni flows.
However, surprisingly, above a certain temperature ($T_C \sim 37\, ^\mathrm{o}$C), a sharp transition into an abrupt spreading motion is observed (Fig.~\ref{fig:MacroDropDynamics}a 6.5--9.2 s, Fig.~\ref{fig:MacroDropDynamics}b).
Shortly thereafter, droplet spreading is accompanied by nucleation and growth of DPnP-rich microdroplets (Fig.~\ref{fig:MacroDropDynamics}a 9.2 s, Supplementary Video~S1).
In contrast, droplets on the partially wetted surfaces phase-separate in their bulk as temperature increases, without any apparent change of droplet radius (Supplementary Video~S2).
This latter scenario is consistent with previous reports on evaporation-driven phase separation (segregation) of binary~\citep{li2018evaporation, kim2018direct} or ternary droplets~\citep{tan2016evaporation, tan2017self} that are subject to contact line pinning.

As a control experiment without phase separation, for instance to clarify the impact of thermal (Marangoni) convection~\citep{ristenpart2007influence, diddens2017Evaporating, gurrala2019evaporation, williams2021spreading}, we replace DPnP with a glycol of similar surface activity, propylene glycol ($\phi = 0.1$, Sigma-Aldrich, $\ge$ 99.5\%)~\citep{cira2015vapour, karpitschka2017marangoni}.
Note that water and propylene glycol are well miscible, meaning that their mixture does not exhibit liquid-liquid phase separation throughout our experimental conditions.
In this case, we only see an enhanced contraction and no spreading motion, during the heating process.

\textit{Impact of phase separation on spreading.---}For pure liquids, the dynamics of advancing contact lines follows the classical Cox-Voinov law~\citep{cox1986dynamics, voinov1976hydrodynamics}
\begin{equation}
\theta_{app}^3 - \theta_{eq}^3 = 9\, \mathrm{Ca}\, \mathrm{ln}\left(\alpha \,\frac{l_o}{l_i}\right),
\label{eq:CVLaw}
\end{equation}
where $\theta_{app}$ and $\theta_{eq}$ denote the dynamical apparent contact angle and the equilibrium contact angle, respectively. $\mathrm{Ca} = \mu U_{CL}/\gamma_{LV}$ is the capillary number, $\mu$ the dynamic viscosity, and $U_{CL}$ the speed of the moving contact line.
$\alpha$ is a nonuniversal numerical constant, and $l_o$ and $l_i$ indicate an outer (macroscopic) and an inner (microscopic) length~\citep{eggers2004characteristic, snoeijer2013moving}. 
Fig.~\ref{fig:MacroDropDynamics}(c) shows the dependence of $\theta_{app}^3 - \theta_{eq}^3$ on $\mathrm{Ca}$ and, on the inset, on $U_{CL}$,  for phase-separating DPnP-water ($\phi = 0.1$, light green and $\phi = 0.15$, dark green), toluene (dark gray) and water (light gray) droplets, respectively.
Here, for the DPnP-water mixtures, toluene and water, $\theta_{eq}$ are $\sim 6.1^\mathrm{o}$ ($\phi = 0.1$), $\sim 6.3^\mathrm{o}$ ($\phi = 0.15$), $\sim 5.5^\mathrm{o}$ and $\sim 2.7^\mathrm{o}$, respectively.
As expected, experimental data of toluene and water collapse onto a master curve (black straight line), following the Cox-Voinov law (Eq.~\ref{eq:CVLaw}).
Surprisingly, we observe a deviation from the Cox-Voinov law for the phase-separating binary droplets, evidencing a direct coupling of phase separation to the spreading process.
We find larger capillary numbers during phase separation, and a power law with an exponent greater than 1 that increases as the glycol-ether mass fraction is increased.
Thus, phase separation accelerates spreading (Fig.~\ref{fig:MacroDropDynamics}c).

Of course, there is no single well-defined capillary velocity, $U_{cap} = \gamma_{LV}/\mu$ for a droplet with ongoing phase separation, since in general, the two phases exhibit different viscosities, and an emulsion may, on top, show non-Newtonian behavior~\citep{derkach2009rheology}.
Nonetheless, it is instructive to non-dimensionalize the contact line velocity $U_{CL}$ with a characteristic value that is representative for the given situation.
We measure surface tensions and viscosities in water-rich ($\phi = 0.1$ and $0.15$) and the corresponding DPnP-rich one-phase regions at the temperature that droplet starts spreading, obtaining for $\phi = 0.1$ capillary velocities $\sim 40$ m/s and $\sim 6.5$ m/s, respectively, and for $\phi = 0.15$ capillary velocities $\sim 18$ m/s and $\sim 4.2$ m/s, respectively. 
In Fig.~\ref{fig:MacroDropDynamics}(c), we use the value for the water-rich phase, which corresponds to the initial condition for the abrupt spreading and the volumetrically dominating phase throughout this process.
The presence of glycol ether-rich microdroplets would increase the apparent viscosity (Supplementary Information, Fig.~S2).
Thus the curves are a lower bound for the actual capillary number.
In the Supplementary Information, we also depict the range of possible capillary numbers (Fig.~S3). 

\begin{figure}
\centering
\includegraphics[width=8.5cm]{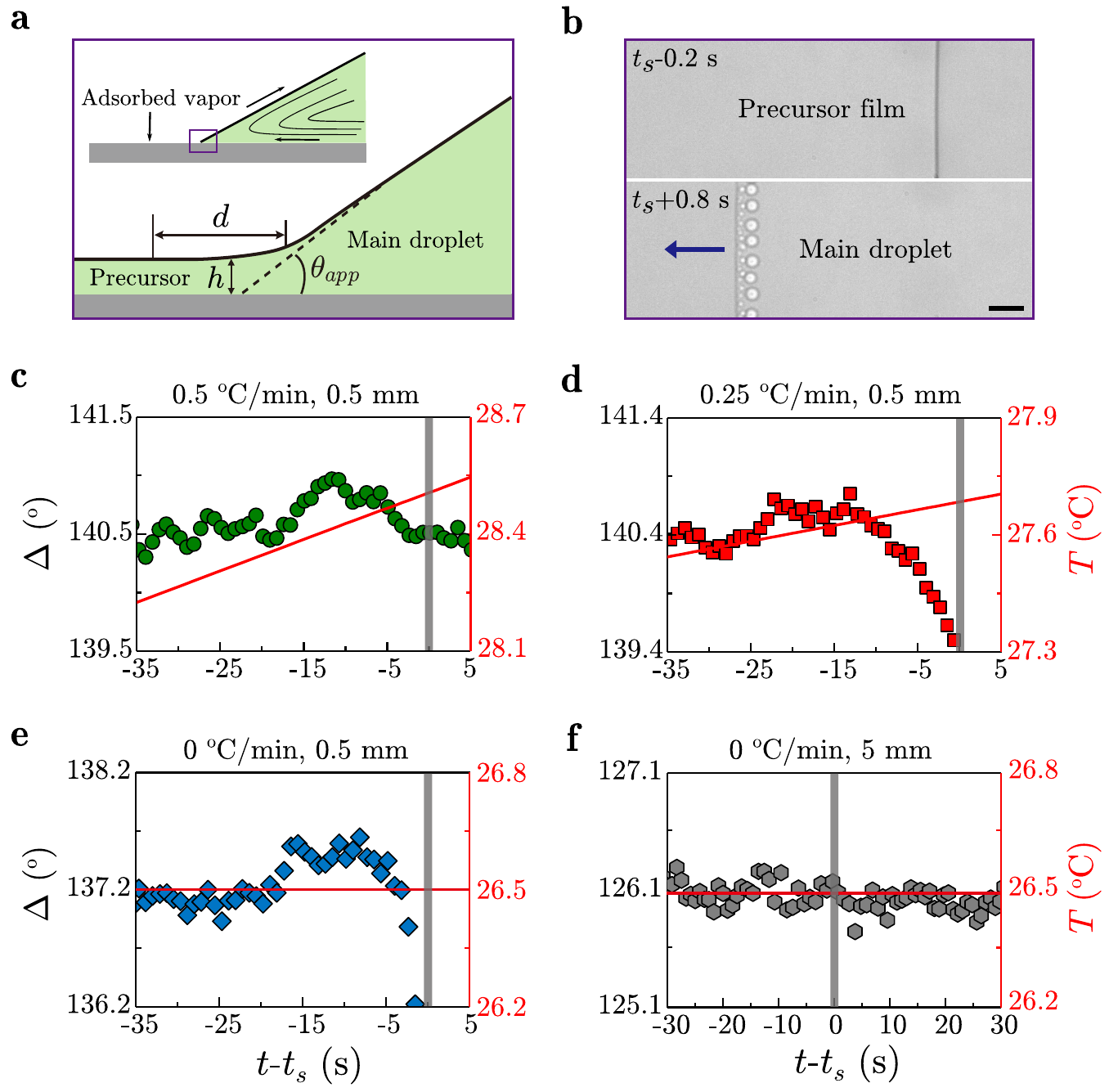}
\caption{Analysis of the precursor film around the main droplet.
(a)~Schematic cross-section of the micro/nano-scopic contact line region. $h$ and $d$ denote film thickness and distance to the contact line, respectively.
(b)~High-resolution images of the contact line region 0.2 s before and 0.8 s after the onset of spreading at $t_s$ (scale bar $\SI{20}{\micro\meter}$).
(c-f)~Ellipsometric angle $\Delta$ (symbols) measured at distance $d$ to the contact line, and temperature $T$ (red lines), vs. $t - t_s$: three different heating rates of $0.5$,  $0.25$, and $0$ $^\mathrm{o}$C/min at $d \sim \SI{0.5}{\milli\meter}$ (c-e, respectively), and $0\, ^\mathrm{o}$C/min at $d \sim \SI{5}{\milli\meter}$ (f).
For constant $T\sim 26.5\, ^\mathrm{o}$C (panels e \& f), phase separation is triggered by selective evaporation. Vertical (grey) lines are a guide to the eye.}
\label{fig:PrecursorDynamics}
\end{figure}

To identify phase separation near the contact line, we further record bright-field images at higher spatial resolution at $\times 40$ magnification (NA 0.60).
As previously observed for pinned droplets~\cite{tan2016evaporation}, microdroplets nucleate and grow at the contact line region (Fig.~\ref{fig:MacroDropDynamics}d and Supplementary Video S3).
However, microdroplets appear only around 0.44 s after the onset of contact line motion (Supplementary Information, Fig.~S4).
We confirm the generality of this phenomenon in our experimental system, using a wide range of heating rates (0.9, $0.3\, ^\mathrm{o}$C/s), different mass fractions of DPnP ($\phi = $ 0.05, 0.15), as well as binary mixtures made up of water and different glycol ethers (TPnP, DPnB) (Supplementary Information, Fig.~S5).
Nucleation is quantified in the images by the average pixel-wise absolute deviation of the intensity in the contact line region from a reference image, a signal that grows rapidly at the point of nucleation (Figs.~S4 and S5).
This suggests physicochemical changes at the contact line or outside the main droplet, i.e., in the adsorbed precursor film, before the visible (macroscopic) phase separation occurs in the main droplet (Fig.~\ref{fig:PrecursorDynamics}a,b).

\textit{Impact of surface forces on phase separation.---}%
Based on above observations (Fig.~\ref{fig:MacroDropDynamics}d), we hypothesize that surface forces drive an earlier change
within the precursor film by promoting liquid-liquid phase separation.
In our system, surface forces are mainly due to van der Waals interactions across the three phases, air/liquid/substrate, which can be quantified in the form of a disjoining pressure~\citep{israelachvili2011intermolecular}
\begin{equation}
\label{eq:disjoining}
\Pi (h) = \frac{A}{6\pi h^3},
\end{equation}
where $A \sim -10^{-20}\SI{}{\joule}$, the Hamaker constant~\citep{israelachvili2011intermolecular}.
Thus, for Eq.~(\ref{eq:disjoining}) to attain values significant to chemical equilibrium, $h \sim \mathcal{O}(\SI{1}{\nano\meter})$ is required, a typical thickness of a precursor film~\citep{bonn2009wetting}.
For complete wetting conditions, surface forces are repulsive, i.e., $\Pi (h)<0$, giving rise to a reduced pressure in the precursor film, $p_{f} = p_\infty + \Pi(h)$ where $p_\infty$ is the ambient pressure.
Pressure is a key factor in chemical equilibrium and well-known to modify the location of phase boundaries~\citep{gibbs1876equilibrium}.
To this end, we conjecture that inside the precursor film, the reduced pressure $p_{f}$ lowers the thermodynamical demixing boundary of the binary mixture, and therefore promotes phase separation.
Literature data for mixtures of water and 2-butoxyethanol shows that the LCST decreases with pressure~\citep{wenzel1980kinetics, compostizo1992thermal}.

To demonstrate the impact of the disjoining pressure on phase separation, we use high-speed \textit{in situ} ellipsometry, which allows for detecting subtle variations of thickness or refractive index in molecularly thin films~\citep{bahr2007surface, popescu2012precursor} (see Methods and Supplementary Information, Fig.~S6).
Figs.~\ref{fig:PrecursorDynamics}(c-f) show the ellipsometric angle $\Delta$ (symbols) as a function of time relative to the onset of spreading, $t - t_s$, at three different heating rates ($0.5$, $0.25$, and $0 \, ^\mathrm{o}$C/min, panels c-e, respectively).
Note that phase separation also occurs at constant temperatures above the LCST, due to selective evaporation (panels e \& f).
For panels c-e, the measurement spot is located at a distance $d \approx 0.5$ mm away from the macroscopic contact line (see Fig.~\ref{fig:PrecursorDynamics}a, and Fig.~S6).
For panel f, the distance is around $\SI{5}{\milli\meter}$.

Long before the onset of spreading ($t - t_s \lesssim \SI{-20}{\second}$), we find a fluctuating $\Delta$.
We attribute this to the fluctuations in the evaporation/condensation equilibrium between the vapor and the hydrophilic surface~\cite{hack2021wetting}, which is also observed when placing a pendant droplet above a fully wetted substrate (Supplementary Information, Fig.~S7).
Around $t-t_s\sim\SI{-20}{\second}$, we observe an abrupt increase of $\Delta$, which is small but distinguishable from noise and reproduced in all repetitions of these experiments.
Another $\sim 5-10$ $\SI{}{\second}$ later, we see a rapid decrease of $\Delta$.
On the contrary, at a large distance to the droplet (comparable to its radius), we do not observe any measurable change in the ellipsometric signal (panel f).

Our results clearly evidence the existence of a composition or morphology variation in the precursor close to the droplet, ahead of any macroscopically visible effect, most probably caused by earlier nucleation in the precursor film.
This variation is sensitive to the distance $d$ from the macroscopic contact line, since the precursor film is, due to its microscopic thickness, always very close to equilibrium with the vapor above it, and the vapor density around an evaporating droplet decays $\sim 1/d$~\cite{deegan2000contact, eggers2010nonlocal}.
Far from the contact line, the less volatile DPnP molecules are outnumbered by the more volatile water molecules, which are abundant in the atmosphere due to the natural humidity.
As such, the phase boundary is hardly ever reached far from the droplet, even though the precursor film becomes much thinner~\cite{novotny1991wetting}.
We note here that no visible increase of $\Delta$ could be observed for large heating rates  $\gtrsim 1.2\, ^\mathrm{o}$C/min (Supplementary Information, Fig.~S8), for which the effect is probably beyond the sampling period ($\sim 0.8$ s) of our ellipsometer.

\begin{figure}
\centering
\includegraphics[width=7.5cm]{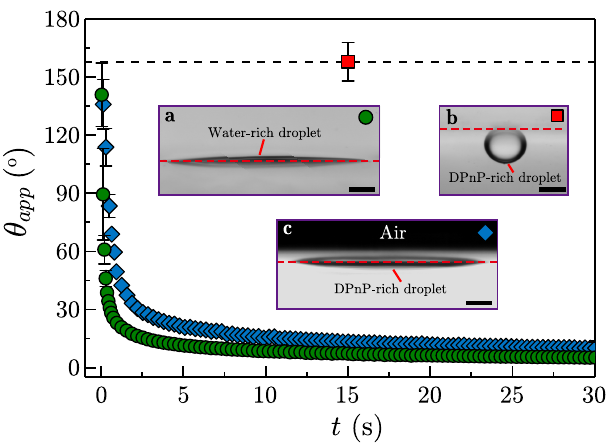}
\caption{Preferential wetting of the two mutually saturated liquid phases at the interface with glass and air: contact angle vs. time (main panel) and side aspects of the immersed droplets (insets).
(a)~Water-rich droplet spreading on hydrophilic glass in a DPnP-rich ambient phase (green circles).
(b)~DPnP-rich droplet dewetted from hydrophilic glass (located above the droplet to have buoyancy pushing the drop against the glass) in a water-rich ambient phase (red square).
(c)~DPnP-rich droplet spreading at the free surface of a water-rich ambient phase (blue diamonds).
Red dashed lines represent the surface location, and scale bars are 0.5 mm.
}
\label{fig:WettabilityTest0}
\end{figure}

\textit{Preferential wetting in the two-phase region.---}%
Finally, to rationalize on which surface (solid-liquid or liquid-vapor) nucleation first emerges, we test the wetting preference of water-rich and glycol ether-rich droplets on either surface.
Here, mutually saturated water-rich and DPnP-rich phases are extracted from bottom and top phases of a well equilibrated DPnP-water mixture at 1:1 mass ratio, respectively (Methods).
A water-rich droplet in a DPnP-rich outer phase (green, Fig.~\ref{fig:WettabilityTest0}a) preferentially wets the clean glass, spreading to small contact angles.
Exchanging droplet and outer phases (red, Fig.~\ref{fig:WettabilityTest0}b), the contact angle remains large, close to 180 degrees.
In both cases, buoyancy is used to push the droplet against the substrate, which is thus located above the droplet in the latter case. 
The opposite behavior is observed at the liquid-air interface, where the DPnP-rich phase spreads along the free surface (blue, Fig.~\ref{fig:WettabilityTest0}c, see also Supplementary Information, Fig.~S9 for additional cases on fully and partially wetted substrates).
These observations suggest that the glycol ether-rich phase nucleates initially at the liquid-air interface, before microdroplets appear in the bulk of the main droplet or at the substrate surface.

Although these observations can not readily be transferred to the precursor region, where the presence of three phases in close proximity leads to strong surface forces that may alter wetting preferences, it renders a ``leaking-out''~\citep{brochard2000wetting} of DPnP rather unlikely.
Instead, an increased water fraction in the precursor film would be expected.
Yet, there the phase change appears well before the spreading or visible microdroplet nucleation.

\textit{Discussion and conclusions.---}In contrast to binary or ternary droplets on partially wetted surfaces with pinned contact lines~\citep{tan2016evaporation, tan2017self, li2018evaporation, kim2018direct, christy2011flow, kim2016controlled, edwards2018density, li2019gravitational, moon2020evaporation, guo2021non}, here we report unexpected spreading of phase-separating binary-mixture droplets on fully wetted surfaces with free contact lines.
Interestingly, we find that Cox-Voinov law is not applicable anymore for such droplets, because phase separation accelerates the speed of moving contact line.
A closer inspection of the contact line region reveals that the nucleation in the main droplet occurs later than the advancing motion of contact line.
We argue that surface forces inside the wetting precursor shift the thermodynamic phase boundary considerably, promoting phase separation well in advance to the observed spreading or bulk droplet behavior, which we verify by ellipsometric measurements.
We demonstrate that the nucleating (glycol ether-rich) phase has a strong wetting preference for the liquid-air interface.
We therefore conclude that, when the binary mixture is pushed into the two-phase region, surface forces facilitate phase separation at the free surface, in nanoscopic proximity to the contact line.
It is this earlier phase separation that drives the contracted droplet away from stationary state and changes the force balance at the three-phase contact line~\citep{cira2015vapour, karpitschka2017marangoni}, thus causing the active spreading.
It is worth noting that no apparent spreading motion can be seen when starting with droplets on the glycol ether-rich side of the miscible region, such as DPnP with $\phi = 0.8$ (see Supplementary Video~S4).
Here, the free surface would already be ether-rich, and no abrupt change in surface energies is expected upon phase separation.
Further, the molecular autophobicity of glucol ethers on glass may prevent the droplet from spreading over its own adsorbed film, causing contact line pinning~\citep{novotny1991wetting, hack2021wetting}.

To summarize, we have demonstrated that the strong coupling between phase separation, moving contact lines, and surface forces results in forced spreading on complete wetting surfaces, well beyond the capillarity-dissipation balance of Cox-Voinov spreading.
Our work shows experimentally the crucial roles of phase separation and surface forces in dynamic multi-phase systems, motivating future studies to reveal the molecular processes and a theoretical understanding of these observations~\citep{thiele2013gradient}.
We expect these findings to enrich also the fundamental understanding of active wetting transitions in tissue morphogenesis~\citep{perez2019active}, phase separation nano-engineering applications~\citep{wasan2003spreading, sadafi2020evaporation, pahlavan2021evaporation}, or liquid-liquid phase separation dynamics in cell biology~\citep{banani2017biomolecular, shin2017liquid}.

\textit{Acknowledgements---}We acknowledge financial support from the Max Planck -- University of Twente Center for Complex Fluid Dynamics. Y.C. acknowledges support through an Alexander von Humboldt Fellowship. We also would like to thank L. D. Rodriguez, A. Barthel, W. Keiderling, K. Hantke, J. Chateau and H. Jeon for assistance with the experiments.

\subsection{Methods}
\textit{Preparation of the substrates.---}Microscopy coverslips (VWR, 24 $\times$ 24 mm, thickness 0.17 mm) or one-side frosted microscopy slides (Corning, 75 $\times$ 25 mm, thickness 0.96--1.06 mm) were treated with either ethanol, plasma, or piranha solutions before use. For ethanol cleaning, the substrates were sonicated in ethanol for 20 min, and then stored in fresh water (``Milli-Q", resistivity 18 M$\Omega$ cm). For plasma cleaning, the substrate was cleaned in an acetone solution in the ultrasonic bath for 15 min,
sonicated in ethanol solution for 15 min, and then rinsed with deionized water, dried in the oven, and finally treated with oxygen plasma (Harrick Plasma) for $\sim 3$ min.
For piranha cleaning, the substrates were treated in piranha solutions (hydrogen peroxide 30\% and sulfuric acid 95\%, VWR, mixture 1:3 by volume) for 20 min. Then, the substrates were rinsed with fresh water for five times, sonicated in hot water ($\sim 80\, ^\mathrm{o}$C) for 10 min, and stored in fresh water. The ethanol and piranha cleaned substrates were used on the day of preparation, and dried with a nitrogen drying gun in a laminar flow hood immediately before each measurement. The plasma-cleaned substrate was used immediately after preparation.

\textit{Preparation of the binary mixtures.---}For binary solutions, we prepared a mixture consisting of water (``Milli-Q" water, resistivity 18 M$\Omega$ cm) and one of the following glycol ethers: di(propylene glycol) propyl ether (DPnP, $\ge$ 98.5\%), tri(propylene glycol) propyl ether (TPnP,  97\%), and di(propylene glycol) butyl ether (DPnB, $\ge$ 98.5\%) or a typical glycol: propylene glycol ($\ge$ 99.5\%). For spreading of pure liquids, we use ``Milli-Q" water and toluene ($\ge$ 99.9\%). All chemicals were purchased from Sigma-Aldrich. The corresponding experimental phase diagrams of water and glycol ether mixture (data from Ref.~\citep{bauduin2004temperature}) are shown in Fig.~S1 (Supplementary Information), and the basic physicochemical parameters of glycol ethers at room temperature ($\sim 25\, ^\mathrm{o}$C) from existing literatures~\citep{staples2002examination, bauduin2004temperature, frank2007separation} are summarized in Table~S1 (Supplementary Information).

\textit{Preparation of two mutually saturated phases.---}The immiscible DPnP-water solution were prepared by first mixing DPnP and water with a weight ratio of 1:1. The well-mixed solution was then centrifuged in a laboratory centrifuge (Centrifuge 5804R, Eppendorf) at 4000 r.p.m for 2 hours and allowed to phase-separate for more than 48 hours. Finally, the two mutually saturated phases, i.e., water-rich and DPnP-rich solutions, were collected from the bottom and top layers of the well equilibrated mixture, respectively.

\textit{Measurements of liquid viscosity, surface tension and contact angle.---}The viscosity $\mu$ of liquid mixture was measured by a temperature-controlled rheometer (Anton Paar MCR 502). The surface tension $\gamma_{LV}$ was measured with a goniometer (DataPhysics OCA 20) using the pendant drop method. For each mixture, at least ten droplets were measured and analyzed to obtain the surface tension, with an average error of 0.16 mN/m. The static apparent contact angle $\theta_{app}$ (Fig.~\ref{fig:PhaseDiagram}e) was measured with this goniometer, using the sessile drop method.

\textit{Observation of the main droplet.---}The recording of main droplet was performed in a custom-built chamber ($\sim$ 10 $\times$ 10 $\times$ 5 cm), mounted on the top of an inverted (epi-fluorescence) microscope (Nikon Eclipse Ti2E).
A computer-controlled heating system was built into the chamber, allowing set a well-defined temperature and heating rate at the substrate.
The heating system composed of a transparent ITO glass (28 $\times$ 28 mm, thickness 0.7 mm, CEC020B, Praezisions Glas \& Optik GmbH), a custom-built PID controller, and a Python-based controlling interface, which was also calibrated by an IR thermal imaging camera (Laserliner).
Droplets composed of mixtures of water (``Milli-Q") and a glycol ether or a propylene glycol, with initial volumes $\Omega=$ 0.5--2 $\SI{}{\micro\liter}$, were gently deposited onto the cleaned microscope coverslip with a glass syringe (Hamilton GasTight).
Afterward, droplet behavior was observed simultaneously by two cameras: one high-speed camera (Phantom VEO 4K-L, 50--500 fps) for the bottom-view recording, and another cmos camera (Point Grey Grasshopper2, 27 fps) attached to a macro lens (Thorlabs Bi-Telecentric lens, 1.0X, W.D. 62.2 mm) with a collimated light source for the side-view recording (also see Fig.~\ref{fig:PhaseDiagram}d)~\citep{ramirez2021taylor}.
The bright-field microscopy was performed with either a $\times 2$ Plan Apo objective for observing the whole droplet or a $\times 40$ (numerical aperture 0.60) Plan Fluor objective for observing the contact line region.
The relative humidity RH and ambient temperature were stable during the experiments, $(30 \pm 5)\%$ and $(21 \pm 1)\, ^\mathrm{o}$C, respectively. 
All images were analyzed by custom-made \textsc{matlab} codes and/or the open-source software \textsc{imagej}.
The dynamical apparent contact angle $\theta_{app}$ (Fig.~\ref{fig:MacroDropDynamics}c) was obtained from the side-view images as $\theta_{app} \simeq 2 h_0/R$, where $h_0$ and $R$ are the maximal height and the foot radius of the droplet, respectively~\citep{ramirez2021taylor}.

\textit{On-site ellipsometric measurements of the precursor film.---}The variation of precursor film was detected in separate experiments using a high-speed phase-modulated ellipsometer (NeHe laser beam, diameter 0.63 mm, wavelength $\lambda = 633$ nm, sampling period $\sim 0.8$ s)~\citep{bahr2007surface}.
To regulate the substrate temperature, the ellipsometer was equipped with a temperature controller (Eurotherm), which was sampled by a thermal sensor.
Furthermore, to minimize the light reflection from the bottom side of glass substrates, one-side frosted microscope slides, instead of coverslips, were applied, and additionally a half part of the substrate were untreated so as to fix the droplet during measurements.
The angle of incidence $\alpha_i$ was adjusted so that the value of ellipsometric angle $\Delta$ was near $135^\mathrm{o}$.
Typically, low heating rates ($\lesssim 1.2\, ^\mathrm{o}$C/min) were applied in order to efficiently capture the fast dynamics of precursor film, and additionally, droplets with large size $\Omega=$ 5--10 $\SI{}{\micro\liter}$ were adopted to reduce the evaporation-induced volume shrinkage.
During ellipsometric measurements, the main droplet was simultaneously recorded by a camera  (Point Grey Grasshopper2, 10 fps) from the top-view, accompanying with a green light ($\lambda = 550$ nm, KL 1500 LCD) as illumination from the side (Supplementary Information, Fig.~S6).
Here, the green light was applied to avoid its interference with the HeNe laser light ($\lambda = 633$ nm).
Under this condition, the value of $\Delta$ was assumed to be most sensitive to the variation inside the precursor film. Measurements were also performed in an atmospheric control chamber to minimize external disturbances in the vapor field due to ambient air currents. 
For all experiments, each measurement begun when the deposited droplet reached a steady state ($\sim$ 1--3 min). During this period, droplets were assumed to form an effective $\theta_{app}$ as well as to develop a stable precursor film. 

\bibliography{mybibfile}

\end{document}